\documentclass[prx,twocolumn, amsmath,amssymb,
reprint,%
author-year,%
author-numerical,%
superscriptaddress
]{revtex4}

\usepackage{graphicx}
\usepackage{bm}
\usepackage{color}
\usepackage{ulem}

\begin{document}


\title{Combined impact of entropy and carrier delocalization on charge transfer exciton dissociation at the donor-acceptor interface}

\author{Shota Ono}
\email{shota_o@gifu-u.ac.jp}
\affiliation{Department of Electrical, Electronic and Computer Engineering, Gifu University, Gifu 501-1193, Japan}
\affiliation{Department of Physics, Graduate School of Engineering, Yokohama National University, Yokohama, 240-8501, Japan}
\author{Kaoru Ohno}
\affiliation{Department of Physics, Graduate School of Engineering, Yokohama National University, Yokohama, 240-8501, Japan}

\begin{abstract}
Several models of the charge transfer exciton (CTE) have been proposed to explain its dissociation at the donor-acceptor (DA) interface. However, the underlying physics is still under debate. Here, we derive temperature ($T$)-dependent tight-binding model for an electron-hole pair at the DA interface. The main finding is the existence of the localization-delocalization transition at a critical $T$, which can explain the CTE dissociation. The present study highlights the combined effect of entropy (finite-$T$) and carrier delocalization in the CTE dissociation. 
\end{abstract}



\maketitle

\section{Introduction}
The origin of exciton dissociation at the donor-accepter (DA) interface is one of the most important problems in the field of organic solar cells. A key to solve the problem is the charge transfer exciton (CTE). Although several models have been proposed to explain the CTE dissociation \cite{few}, there is no consensus on what physics is mainly involved. The origins derived from the previous models \cite{arkhipov,wiemer,peumans,rubel,deibel,nenashev,schwarz,raos,ono,clarke,gregg,gao,monahan} are listed in Table \ref{tab:previous}; dark dipoles \cite{arkhipov,wiemer,schwarz}, disorder \cite{peumans,rubel,nenashev,raos}, carrier delocalization \cite{deibel,nenashev,raos,ono}, and light effective mass \cite{arkhipov,wiemer,schwarz,ono}, and so forth. In recent years, the entropy effect on the CTE dissociation at the DA interface \cite{clarke,gregg} has been a hot topic. Recent experiments seem to support this view \cite{gao,monahan}. However, the entropy effect has not been studied in terms of the quantum mechanical description where both the electron and the hole are treated by delocalized carriers. 

\begin{figure}[t]
\center
\includegraphics[scale=0.4,clip]{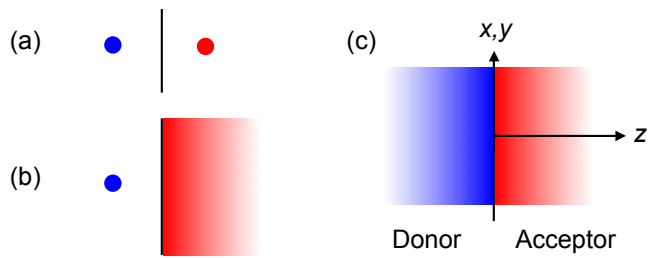}
\caption{\label{fig:treat} (Color online) Schematic illustration of the CTE at DA interface: (a) Localized hole and localized electron, (b) localized hole and delocalized electron, and (c) delocalized hole and delocalized electron.}
\end{figure}

In order to emphasize the importance of the delocalized carrier treatment (quantum mechanical description) for the CTE dissociation, it is useful to introduce the models developed so far. In earliest study, Arkhipov et al. have proposed a dark dipole model to explain the exciton dissociation \cite{arkhipov}. The model considers localized hole or electron [see Fig.~\ref{fig:treat}(a)], one of which goes away from the interface as a result of competing effect between dark dipoles at the DA interface and zero-point oscillations. Based on or inspired by this model, several models have been developed to explain the CTE properties \cite{wiemer,peumans,rubel,deibel,nenashev,schwarz,raos,ono,clarke,gregg,gao,monahan}. In particular, role of delocalization of the electron (or hole) [see Fig.~\ref{fig:treat}(b)] has been pointed out by Nenashev et al \cite{nenashev}. They have studied the effect of charge delocalization along the polymer chains and studied the exciton dissociation rate as a function of applied electric field by using the Miller-Abrahams expression for the hopping rate \cite{miller} and the dissociation probability formula for one-dimensional lattices \cite{rubel}. Recently, combined model of both Arkhipov and Nenashev has been used to explain the field-dependent photocurrent of polymer/C$_{60}$ cells \cite{schwarz}. This study has shown the importance of the dissociation at localized acceptor sites as well as at delocalized donor sites, which indicates the limitation of the localized carrier approximation. Effect of delocalization of both carriers [see Fig.~\ref{fig:treat}(c)] has been studied by Raos et al. using a coarse-grained quantum chemical model \cite{raos}. Within the tight-binding (TB) approximation, they have shown that the sites where charge concentrates are not necessarily those just next to the DA interface, and this holds even in the ground state if diagonal and/or off-diagonal disorder exists. 

In this paper, we propose to establish a quantum mechanical model, a temperature ($T$)-dependent TB model for an Electron-Hole (EH) pair, for studying the entropy effect on the CTE dissociation at the DA interface. Our model is distinct from others in that the entropy as well as the carrier delocalization effect is taken into account. Using this model, we reveal the origin of the CTE dissociation at the DA interface. The main finding of this paper is the existence of the localization-delocalization transition at a critical $T$, which could be considered as an origin of the CTE dissociation. The transition can be observed only when the carrier delocalization treatment is employed. In this way, the CTE dissociation can be successfully explained by a combination of quantum mechanics and thermodynamics. Our work is the first step for understanding the CTE dissociation observed at various DA interface in a unified manner.

The remainder of the paper is organized as follows. In Sec.~\ref{sec:TBA}, we derive the TB model for the EH pair in thermal equilibrium at finite $T$, which allows us to study the finite-$T$ CTE properties. The numerical results are shown in Sec.~\ref{sec:CTED}. In particular, a mechanism of the localization-delocalization transition is explained in Sec.~\ref{sec:mechanism}. Several effects, such as the hopping integrals and disorder, on the transition temperature are investigated in Secs.~\ref{sec:hopping} and \ref{sec:disorder}, respectively. Comparison to the experiments and a possible scenario how the CTE dissociates are discussed in Sec.~\ref{sec:discussion}. Conclusions are presented in Sec.~\ref{sec:Conc}.

\begin{table*}[tpb]
\begin{center}
\caption{The CTE models, origins of the CTE dissociation, and carrier treatment. L/L [Fig.~\ref{fig:treat}(a)] and D/D [Fig.~\ref{fig:treat}(c)] indicate that both electron and hole are treated by localized and delocalized carriers, respectively. L/D [Fig.~\ref{fig:treat}(b)] indicates that one of the carriers is treated by localized carrier, while another is treated by delocalized one.}
{
\begin{tabular}{lll}\hline\hline
Models      & Origin & Carriers  \\ \hline
1D chains \cite{arkhipov,wiemer}  & Dark dipoles, Light mass & L/L   \\ \hline
3D classical TB \cite{peumans}  & Disorder  & L/L    \\ \hline
1D chains \cite{rubel}  & Disorder  & L/L    \\ \hline
2D Molecules \cite{deibel}  & Carrier delocalization   & L/D   \\ \hline
Quasi-1D chains \cite{nenashev}  & Disorder, Carrier delocalization  & L/D   \\ \hline
Quasi-1D chains \cite{schwarz}  & Dark dipoles, Light mass  & L/D    \\ \hline
3D TB \cite{clarke,gregg,gao}  & Entropy   & L/D   \\ \hline
3D Hydrogen \cite{monahan}  & Entropy   & L/D   \\ \hline
2D TB \cite{raos}  & Disorder, Carrier delocalization  & D/D    \\ \hline
3D Hydrogen \cite{ono} & Inversion symmetry breakdown, Light mass, Carrier localization  & D/D  \\ \hline
3D TB (Present work) & Entropy, Carrier delocalization & D/D  \\ \hline
\hline
\end{tabular}
}
\label{tab:previous}
\end{center}
\end{table*}

\section{Formulation: Tight-binding approximation}
\label{sec:TBA}
\begin{figure}[t]
\center
\includegraphics[scale=0.6,clip]{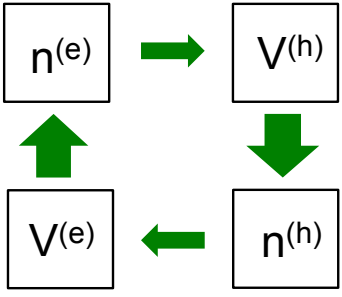}
\caption{\label{fig:DA} (Color online) Schematic representation of our TB model for a EH pair. The thick arrows denote the solution of Eq.~(\ref{eq:sch}). The thin arrows denote Eqs.~(\ref{eq:Vele}) or (\ref{eq:Vhole}).}
\end{figure}

We derive the self-consistent equations for the electron and the hole at the DA interface within the TB approximation. When only one photon is absorbed near the DA interface, an EH pair is created. In this case, we can ignore the electron-electron and hole-hole interaction energy. The Schr\"{o}dinger equation for electron ($i={\rm e}$) and hole ($i = {\rm h}$) in the effective medium is given by
\begin{eqnarray}
 {\cal H}^{(i)}\vert \phi_{\alpha}^{(i)} \rangle 
 = \varepsilon_{\alpha}^{(i)} \vert \phi_{\alpha}^{(i)} \rangle,
 \label{eq:sch}
\end{eqnarray}
where $\phi_{\alpha}^{(i)}$ and $\varepsilon_{\alpha}^{(i)}$ are, respectively, the eigenfunction and eigenenergy with a quantum number $\alpha$ for a particle $i$. By discretizing the space variables and considering the cubic lattice, the TB Hamiltonian for $i={\rm e}$ and ${\rm h}$ is given as
\begin{eqnarray}
 {\cal H}^{(i)} &=& -\sum_{\bm{p},\bm{p'}} t_{\bm{p},\bm{p'}}^{(i)}
 \vert \bm{p} \rangle \langle \bm{p'} \vert
  + \sum_{\bm{p}} V_{\bm{p}}^{(i)}\vert \bm{p} \rangle \langle \bm{p} \vert, 
\end{eqnarray}
where the first and second terms in the right hand side are the kinetic energy and the effective potential energy, respectively. $t_{\bm{p},\bm{p'}}^{(i)}$ ($i=$e or h) is the hopping parameter. The sets of integers $\bm{p} =(p_x,p_y,p_z)$ denotes the electron and hole positions. The magnitude of the hopping parameters is, in principle, determined by the electronic band structure of molecular crystals of donor and acceptor. Thus the hopping parameter may have long-range part and show directional anisotropy. These are originated from the periodic potential part of the external potential, i.e., the nuclear potential of the molecular crystals. The sum of the remaining external potential part $w_{\bm{p}}^{(i)}$ and the EH interaction energies determines the effective potential energy $V_{\bm{p}}^{(i)}$ expressed by 
\begin{eqnarray}
 V_{\bm{p}}^{({\rm e})} &=& w_{\bm{p}}^{({\rm e})} - U_0 \sum_{\bm{p}'} 
 \frac{n^{({\rm h})}_{\bm{p}'}}{\vert \bm{p} - \bm{p}' \vert}, 
 \label{eq:Vele} \\
 V_{\bm{p}}^{({\rm h})} &=& w_{\bm{p}}^{({\rm h})} - U_0 \sum_{\bm{p}'} 
 \frac{n^{({\rm e})}_{\bm{p}'}}{\vert \bm{p} - \bm{p}'\vert},
 \label{eq:Vhole}
\end{eqnarray}
where $U_0 = e^2/(4\pi \varepsilon d)$ is the strength of the EH interaction energy. $d$ is the bond length between sites and $\varepsilon$ is the dielectric constant. The potentials $w_{\bm{p}}^{(i)}$ play an important role in describing the DA interface, which will be discussed below. The $T$-dependent electron and hole densities in Eqs.~(\ref{eq:Vele}) and (\ref{eq:Vhole}) are defined as
\begin{eqnarray}
 n^{(i)}_{\bm{p}} &=& \sum_{\alpha}^{\rm all}  f_{\alpha}^{(i)}
 \vert \langle \bm{p} \vert \phi_{\alpha}^{(i)} \rangle \vert^2, 
  \nonumber\\
 f_{\alpha}^{(i)} &=& \left[ e^{(\varepsilon_{\alpha}^{(i)}-\mu^{(i)})/(k_{\rm B}T)} + 1\right]^{-1},
 \label{eq:rhoele}
\end{eqnarray}
where $i = {\rm e \ or \ h}$. $f_{\alpha}^{(i)}$ and $k_{\rm B}$ are the Fermi distribution function and the Boltzmann constant, respectively. $\alpha$ in the sum runs over all eigenstates. Each density is normalized to unity, i.e., $\sum_{\bm{p}} n^{({\rm e})}_{\bm{p}} = \sum_{\bm{q}} n^{({\rm h})}_{\bm{q}} = 1$ with the use of the chemical potential $\mu^{(i)}$. Since the effective potential for electron given by $V_{\bm{p}}^{({\rm e})}$ depends on the hole density given by $n^{({\rm h})}_{\bm{p}}$ and vice versa, the Schr\"{o}dinger equation in Eq.~(\ref{eq:sch}) for electron and hole should be solved self-consistently for each $T$. The mathematical structure of our model is illustrated in Fig.~\ref{fig:DA}: the electron density $n^{({\rm e})}_{\bm{p}}$, the on-site energy for hole $V^{({\rm h})}_{\bm{p}}$, the hole density $n^{({\rm h})}_{\bm{p}}$, and the on-site energy for electron $V^{({\rm e})}_{\bm{p}}$ are determined by $V^{({\rm e})}_{\bm{p}}$, $n^{({\rm e})}_{\bm{p}}$, $V^{({\rm h})}_{\bm{p}}$, and $n^{({\rm h})}_{\bm{p}}$, respectively. This loop forms the self-consistent equation.

The free energy (the sum of the kinetic energy, the potential energy, and the entropic energy) in the present TB approximation is expressed as
\begin{eqnarray}
 \Omega &=& \sum_{i= {\rm e, h}} \Omega_{0}^{(i)}
 + U_0 \sum_{\bm{p}}  \sum_{\bm{p}'} 
 \frac{n^{({\rm e})}_{\bm{p}} n^{({\rm h})}_{\bm{p}'}}{\vert \bm{p} - \bm{p}' \vert}
 \nonumber\\
 \Omega_{0}^{(i)} &=& 
 \sum_{\alpha} \varepsilon_{\alpha}^{(i)}f_{\alpha}^{(i)}
 -Ts^{(i)}
 \label{eq:OmegaTB}
\end{eqnarray}
with the entropic contribution
\begin{eqnarray}
s^{(i)} = - k_{\rm B}\sum_{\alpha} 
\left[
f_{\alpha}^{(i)} {\rm ln} f_{\alpha}^{(i)}
+ (1- f_{\alpha}^{(i)}) {\rm ln} (1 - f_{\alpha}^{(i)}) 
\right].
\label{eq:Ss}
\end{eqnarray}
Equation (\ref{eq:OmegaTB}) is an approximated version of  the free energy in two-component many-body systems. The derivation of the free energy is similar to that in finite-$T$ \cite{mermin} and two-component \cite{chakra,boronski} Density Functional Theory (DFT). Such a theory will be presented elsewhere \cite{ono_ohno}. For later use, we decompose $\Omega$ into the internal energy $U_{\rm int}$ 
\begin{eqnarray}
 U_{\rm int} &=& E_{\rm band} + E_{\rm int}, \nonumber\\
 E_{\rm band} &=& \sum_{i= {\rm e, h}}  \sum_{\alpha} \varepsilon_{\alpha}^{(i)}f_{\alpha}^{(i)},
 \nonumber\\
 E_{\rm int} &=& U_0 \sum_{\bm{p}}  \sum_{\bm{p}'} 
 \frac{n^{({\rm e})}_{\bm{p}} n^{({\rm h})}_{\bm{p}'}}{\vert \bm{p} - \bm{p}' \vert},
 \label{eq:Uint}
\end{eqnarray}
where $E_{\rm band}$ and $E_{\rm int}$ are the band energy and interaction energy between an electron and a hole, respectively, and the entropic energy
\begin{eqnarray}
 -TS = -T\sum_{i= {\rm e, h}} s^{(i)}.
 \label{eq:TS}
\end{eqnarray}
In the present study, we use a hopping parameter $t_0$ as an energy units. 

\begin{figure}[t]
\center
\includegraphics[scale=0.35,clip]{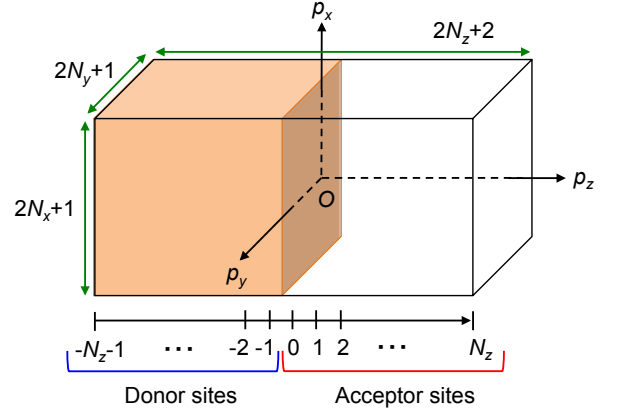}
\caption{\label{fig:DA_TB} (Color online) Simple cubic lattice for the DA interface model. The donor and acceptor regions are $-N_z -1 \le p_z \le -1$ and $0 \le p_z \le N_z$, respectively. The total number of sites is $(2N_x+1)(2N_y+1)(2N_z+2)$. }
\end{figure}

\section{Charge transfer exciton dissociation}
\label{sec:CTED}
We consider an electron and a hole in a simple cubic lattice shown in Fig.~\ref{fig:DA_TB}. The motion of the charged carriers is confined to the region of $-N_x \le p_x \le N_x$, $-N_y \le p_y \le N_y$, and $-N_z - 1\le p_z \le N_z$ by infinite potential barriers outside the boundary. The total number of the sites is $(2N_x+1)(2N_y+1)(2N_z+2)$. We assume that the donor and acceptor regions are confined to $-N_z - 1\le p_z \le -1$ and $0 \le p_z \le N_z$, respectively, so that there is a DA interface between $p_z = 0$ and $-1$ shown in Fig.~\ref{fig:DA_TB}. The DA interface can be described by finite potentials $w_{\bm{p}}^{({\rm e})}$ and $w_{\bm{p}}^{({\rm h})}$ in Eqs.~(\ref{eq:Vele}) and (\ref{eq:Vhole}), respectively. The potentials are modeled as
\begin{eqnarray}
 w_{\bm{p}}^{({\rm e})} = 
 \begin{cases}
  w_0 & p_z \le -1 \\
 0 & p_z\ge 0 
 \end{cases},
 \ \ 
  w_{\bm{p}}^{({\rm h})} = 
 \begin{cases}
 0 & p_z \le -1 \\
 w_0 & p_z \ge 0
 \end{cases},
 \label{eq:posipot}
\end{eqnarray}
where $w_0$ determines the strength of the potential well that would separate the carriers into donor or acceptor region. In contrast, the Coulomb potential energy parameter $U_0$ in Eqs.~(\ref{eq:Vele}) and (\ref{eq:Vhole}) determines the attractive interaction that would yield the exciton states. In the limit of $w_0/U_0 \rightarrow \infty$, the electron (hole) does not penetrate the donor (accepter) region \cite{monahan}. In this case, there is no overlap between electron and hole distribution, implying that there is no EH correlation. As $w_0$ decreases, the charge transfer occurs and recovers the EH correlation. In the limit of $w_0/U_0 \rightarrow 0$, the electron and hole occupy the same site due to the attractive Coulomb interaction, which yields the Frenkel exciton. 

To study the CTE dissociation, it is useful to investigate the real-space density distribution along the direction normal to the DA interface. We define the $p_z$-dependence of the electron and hole probability density $Q_{\rm tot}^{(i)}$ with $i = {\rm e \ or \ h}$ as
\begin{eqnarray}
 Q_{\rm tot}^{(i)}(p_z) &=& \sum_{p_x=-N_x}^{N_x}\sum_{p_y=-N_y}^{N_y} n^{(i)}_{\bm{p}}.
 \label{eq:eprob}
\end{eqnarray}
If both of the electron and the hole are localized around the DA interface, the CTE is formed. On the other hand, if these densities are delocalized over the sites, the CTE is dissociated. The CTE properties at the DA interface depend on various quantities: $k_{\rm B}T$, $t_{\bm{p},\bm{p'}}^{(i)}$, and $w_{\bm{p}}^{(i)}$ with $i = {\rm e \ or \ h}$. These effects are investigated below.


\begin{figure}[t]
\center
\includegraphics[scale=0.5,clip]{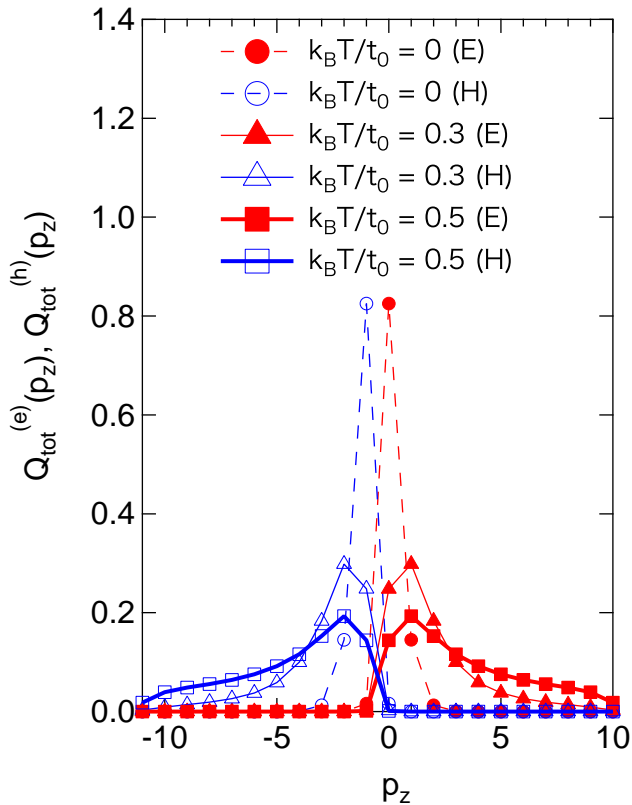}
\caption{\label{fig:prob} (Color online) The $p_z$ dependence of $Q_{\rm tot}^{({\rm e})}$ (filled) and $Q_{\rm tot}^{({\rm h})}$ (open) given by Eq.~(\ref{eq:eprob}) for $k_{\rm B}T/t_0 = 0$ (circle), $0.3$ (triangle), and $0.5$ (square). The values of $U_0/t_0$ and $w_0/t_0$ are set to 10.}
\end{figure}

\subsection{Dissociation mechanism}
\label{sec:mechanism}
We first investigate the finite-$T$ effect on the CTE properties and demonstrate the CTE dissociation at higher $T$. Given the TB model, we may employ various approximations for the hopping integrals depending on the molecule configurations near the DA interface. As the simplest case, we study the TB model with the nearest-neighbor hopping parameters only,
\begin{eqnarray}
 t_{\bm{p},\bm{p}+\bm{R}_{1i}}^{({\rm e})} =
 t_{\bm{p},\bm{p}+\bm{R}_{1i}}^{({\rm h})} = t_0 \ (i = x,y,z),
\end{eqnarray}
where $\bm{R}_{c x} = (\pm c,0,0)$, $\bm{R}_{c y} = (0,\pm c,0)$, and $\bm{R}_{c z} = (0,0,\pm c)$ with a positive integer $c$. 

Let us start with the situation that the electron and hole are strongly localized at, respectively, acceptor and donor site near the DA interface at lower $T$. To investigate it, the parameters $w_0$ in Eq.~(\ref{eq:posipot}) and $U_0$ are both set at $10t_0$. We calculated $\Omega(T=0)$ given in Eq.~(\ref{eq:OmegaTB}) for the case of $(N_x, N_y, N_z) = (M,M,10)$ with a positive integer $M$. We checked the convergence of the value of $\Omega(T=0)$ with respect to $M$ and found that the values for $M=5$ and $M=6$ are the same with an error of less than $0.00001t_0$. For all calculations discussed in this paper, the sizes of $(N_x, N_y, N_z)$ are set to $(5,5,10)$. Results of this model may serve as a reference for those of more complex models below.

\subsubsection{Anomalies at a critical $T$}
Figure \ref{fig:prob} shows the $p_z$ dependence of $Q_{\rm tot}^{({\rm e})}(p_z)$ and $Q_{\rm tot}^{({\rm h})}(p_z)$ in Eq.~(\ref{eq:eprob}) for $k_{\rm B}T/t_0=0, 0.3$, and $0.5$. At zero $T$, $Q_{\rm tot}^{({\rm e})}$ ($Q_{\rm tot}^{({\rm h})}$) has the maximum value of 0.8 at $p_z=0$ ($p_z=-1$) and decays within a few positive (negative) $p_z$s. As $T$ increases, the $p_z$ dependence of $Q_{\rm tot}^{({\rm e})}$ and $Q_{\rm tot}^{({\rm h})}$ changes dramatically at around $k_{\rm B}T/t_0\simeq 0.3$: The value of $Q_{\rm tot}^{({\rm e})}$ and $Q_{\rm tot}^{({\rm h})}$ have the maximum of $\sim 0.3$ at the sites away from those just next to the interface, i.e., $p_z=1$ and $p_z = -2$, respectively, and are averaged out over all $p_z$, which clearly indicates the CTE dissociation. 

\begin{figure}[t]
\center
\includegraphics[scale=0.45,clip]{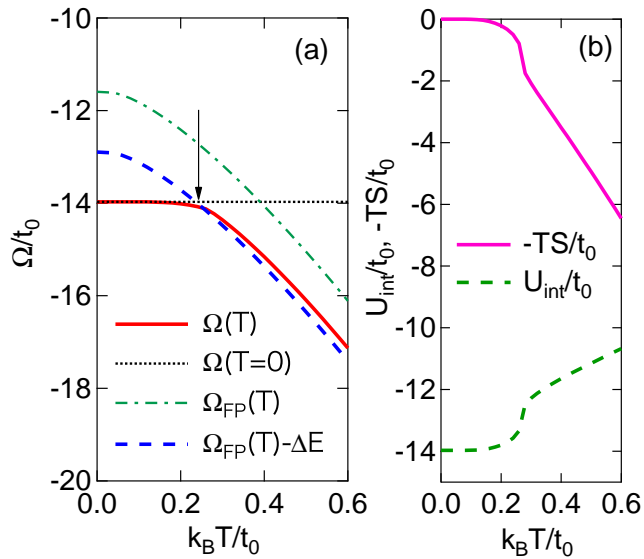}
\caption{\label{fig:free_energy} (Color online) (a) The $T$ dependence of the free energy $\Omega(T)$ (red solid) given by Eq.~(\ref{eq:OmegaTB}). The crossing point between the free-particle free energy $\Omega_{\rm FP}(T)-\Delta E$ (blue dashed) and the straight line $\Omega(T=0)$ (black dotted) is indicated by an arrow. (b) $U_{\rm int}$ in Eq.~(\ref{eq:Uint}) and $-TS$ in Eq.~(\ref{eq:TS}) as a function of $T$. }
\end{figure}

The localization-delocalization transition observed in Fig.~\ref{fig:prob} is related to the free energy anomaly. Figure \ref{fig:free_energy}(a) shows $\Omega$ in Eq.~(\ref{eq:OmegaTB}) as a function of $T$ (red solid). The anomaly in $\Omega$ is observed at a critical temperature $k_{\rm B}T_{\rm c}/t_0\simeq 0.27$. $\Omega$ is almost independent of $T$ below $T_{\rm c}$, while $\Omega$ decreases monotonically with increasing $T$ above $T_{\rm c}$. This localization-delocalization transition across $T_{\rm c}$ is the first finding in this paper. The mechanism of the transition will be shown below. 

\subsubsection{Internal energy and entropy}
To better understand the transition across $T_{\rm c}$ we analyze the $T$ dependence of the free energies. Figure \ref{fig:free_energy}(b) shows the $T$ dependence of the internal energy $U_{\rm int} = E_{\rm band} + E_{\rm int}$ defined as Eq.~(\ref{eq:Uint}) and the entropy $-TS$ defined as Eq.~(\ref{eq:TS}). Similar anomalies are also observed in the $T$ dependence of $U_{\rm int}$ and $-TS$: $U_{\rm int}$ and $-TS$ jump at $T=T_{\rm c}$, below which $U_{\rm int}$ and $-TS$ are almost independent of $T$, and above which $U_{\rm int}$ and $-TS$ increases and decreases, respectively. Since $S\simeq 0$ below $T_{\rm c}$, $\Omega$ is dominated by the contribution from $U_{\rm int}$. On the other hand, $\Omega$ is dominated by the entropy contribution above $T_{\rm c}$. 

\begin{figure}[t]
\center
\includegraphics[scale=0.45,clip]{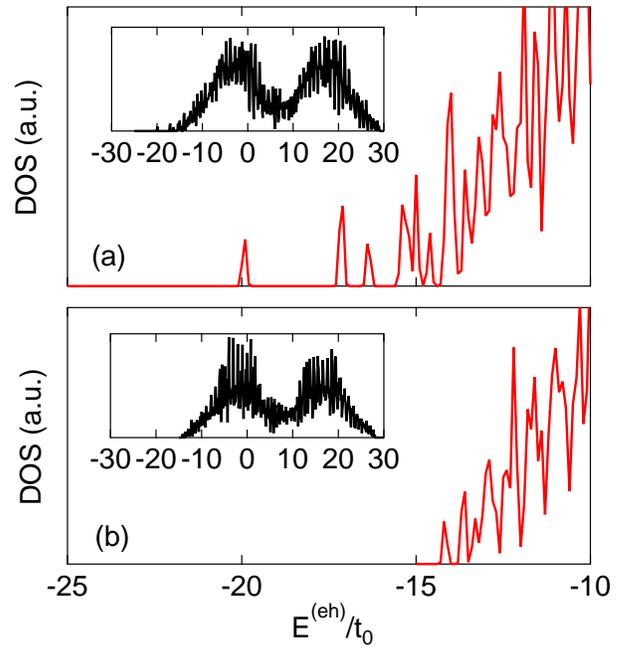}
\caption{\label{fig:dos} (Color online) Electron-hole DOS for (a) $k_{\rm B}T/t_0=$ 0 and (b) 0.6 from $E^{\rm (eh)}/t_0 = -25$ to $-10$. The whole DOS is shown in the inset.}
\end{figure}

\begin{figure}[t]
\center
\includegraphics[scale=0.45,clip]{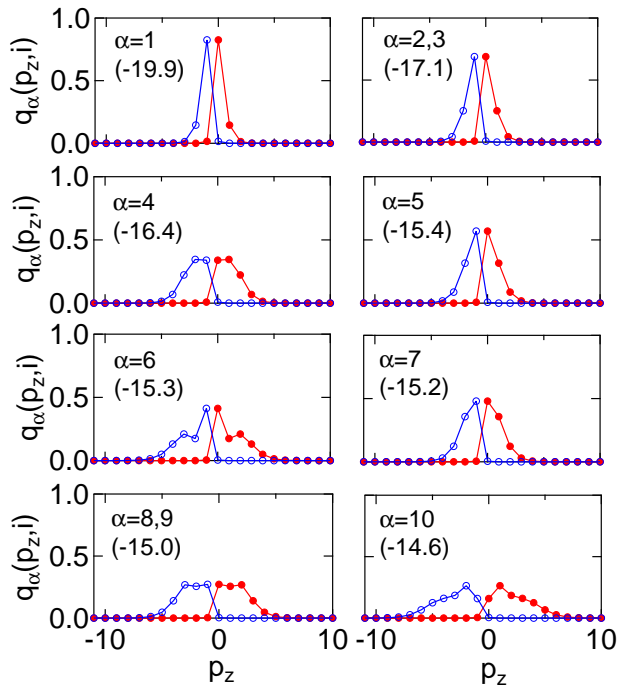}
\caption{\label{fig:charges} (Color online) The $p_z$ dependence of electron (filled circle) and hole (open circle) density from the first to 10th eigenstate at $T=0$. The eigenenergie $E_{\alpha}^{\rm (eh)}$ is also shown in units of $t_0$.}
\end{figure}

\begin{figure}[t]
\center
\includegraphics[scale=0.5,clip]{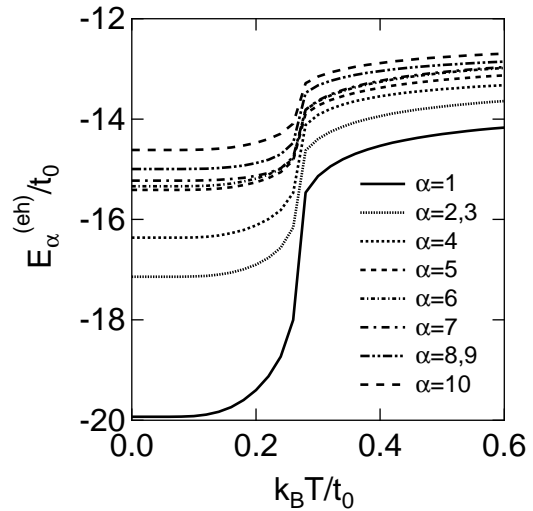}
\caption{\label{fig:Eeh} (Color online) $E_{\alpha}^{\rm (eh)}$ as a function of $T$ for $\alpha=1$--10. The values of $U_0/t_0$ and $w_0/t_0$ are set to 10.}
\end{figure}

\begin{figure}[t]
\center
\includegraphics[scale=0.5,clip]{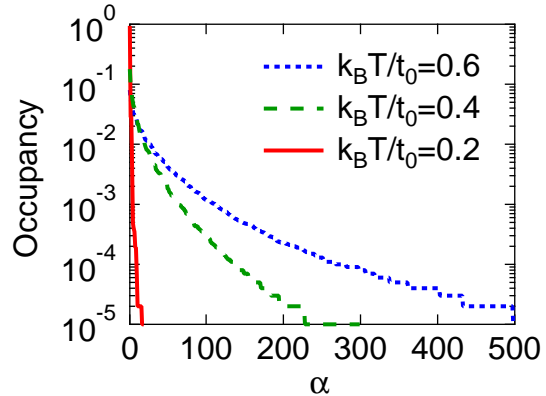}
\caption{\label{fig:Occ} (Color online) The occupation number of the eigenstate with $\alpha$ from 1 to 500 for $k_{\rm B}T/t_0=0.2$, 0.4, and 0.6.}
\end{figure}

The $T$ dependence of $U_{\rm int}$ and $-TS$ at low and high $T$ limit can be understood through the EH eigenstates analysis. To study this, we show the EH density-of-states (DOS) (Fig.~\ref{fig:dos}), the charge distribution (Fig.~\ref{fig:charges}), the eigenenergies (Fig.~\ref{fig:Eeh}), and the occupancy of the eigenstates (Fig.~\ref{fig:Occ}). We start with the EH DOS of $k_{\rm B}T/t_0=0$ and $0.6$, shown in the inset of Fig.~\ref{fig:dos}(a) and \ref{fig:dos}(b), respectively. We define the EH energy as
\begin{eqnarray}
E_{\alpha}^{\rm (eh)} = \varepsilon_{\alpha}^{({\rm e})} + \varepsilon_{\alpha}^{({\rm h})}.
\end{eqnarray}
To draw the DOS, we assumed that each peak is broadened by a Gaussian function with a finite width of $0.05t_0$. Many eigenstates exist from $E_{\alpha}^{\rm (eh)}/t_0 = -15$ to $30$ for both $T$. These are delocalized states. The interesting fact is that the isolated peaks such as $E_{\alpha}^{\rm (eh)}/t_0 = -19.9, -17.1$, and $-16.4$ at lower energies are observed for lower $T$ shown in Fig.~\ref{fig:dos}(a), while no isolated peak is observed for higher $T$ shown in Fig.~\ref{fig:dos}(b). The corresponding probability densities $q_{\alpha}(p_z,i) = \sum_{p_x,p_y}\vert \langle \bm{p} \vert \phi_{\alpha}^{(i)}\rangle \vert^2 $ $(i = {\rm e \ or \ h})$ for the $\alpha$th eigenstate $(\alpha = 1$--$10)$ are shown in Fig.~\ref{fig:charges}. The value of $E_{\alpha}^{\rm (eh)}/t_0$ is also shown. The EH distribution of $\alpha=1$ is exactly the same as $Q_{\rm tot}^{(i)}(p_z)$ at $k_{\rm B}T/t_0=0$ shown in Fig.~\ref{fig:prob} because at zero $T$ the electron and hole occupy the lowest energy level only (i.e., the ground state). As $\alpha$ increases, the distribution of the electron and hole densities are gradually extended into acceptor and donor region, respectively. This shows that the peaks at lower and higher $E_{\alpha}^{\rm (eh)}$ in the EH DOS are originated from the strongly bounded CTE and delocalized EH pair (or weakly bounded CTE), respectively.

One can expect that the peak energies will shift by the finite-$T$ effect. This is true if one calculates the eigenenergy by solving Eq.~(\ref{eq:sch}) self-consistently at each $T$. Since the effective potential $V_{\bm{p}}^{(i)}$ is a functional of the $T$ dependent density, the eigenenergies of the EH states are a functional of the EH density. Hence, the eigenenergies are a function of $T$. Furthermore, one can expect that $E_{\alpha}^{\rm (eh)}$ increases with increasing $T$. The reason is as follows: (i) At finite-$T$, the electron and hole accommodate the excited states as well as the ground state; (ii) Each carrier density would be spatially extended since the total carrier density is a sum of all the probability density of the eigenstates weighted by the Fermi distribution function [see Eq.~(\ref{eq:rhoele})]; (iii) The spatial extent of the charge density reduces the strength of the attractive Coulomb interaction [see Eqs.~(\ref{eq:Vele}) and (\ref{eq:Vhole})], leading to a decrease in the binding energy; and thus $E_{\alpha}^{\rm (eh)}$ increases. 

Figure~\ref{fig:Eeh} shows the lowest 10 eigenenergies $E_{\alpha}^{\rm (eh)}$ as a function of $T$. Indeed, all $E_{\alpha}^{\rm (eh)}$s increase monotonically with increasing $T$. The drastic increase in $E_{\alpha}^{\rm (eh)}$ is observed at $k_{\rm B}T_{\rm c}/t_0=0.27$. The most important fact is that the energy level spacing is quantitatively different between below and above $T_{\rm c}$: The spaces are relatively large (about 2.8$t_0$, 0.8$t_0$, and $t_0$) at lower energy below $T_{\rm c}$, while above $T_{\rm c}$ those are less than $0.5t_0$. Figure~\ref{fig:Occ} shows the occupancy of $\alpha$th state up to $\alpha = 500$ for $k_{\rm B}T/t_0=0.2, 0.4$, and 0.6. The occupation number of the eigenstate with $\alpha =1$ is still over 0.9 at $k_{\rm B}T/t_0=0.2$. Above $T=T_{\rm c}$, the partial occupancy of higher energy levels increases with increasing $T$ due to the small space of the energy levels shown in Fig.~\ref{fig:Eeh}. 

Using these results (Figs.~\ref{fig:dos}--\ref{fig:Occ}), we can interpret the $T$ dependence of $U_{\rm int}$ and $-TS$ at low and high $T$ limit. As expected from Fig.~\ref{fig:Eeh}, the increase in $E_{\alpha}^{\rm (eh)}$ below $T_{\rm c}$ leads to an increase in $E_{\rm band}$ with increasing $T$. $E_{\rm int}$, in turn, decreases due to the spatial extent of the charges shown in Fig.~\ref{fig:charges}, which partly cancels the increase in $E_{\rm band}$. Thus $U_{\rm int}$ is almost independent of $T$ at lower $T$. Simultaneously, $-TS$ is almost zero at lower $T$ from Eq.~(\ref{eq:Ss}) and Fig.~\ref{fig:Occ}. Since the lowest energy state is energetically isolated shown in Figs.~\ref{fig:dos}(a) and \ref{fig:Eeh}, $-TS$ is also independent of $T$. At higher $T$, as is clear from Eq.~(\ref{eq:Ss}) and Fig.~\ref{fig:Occ}, $-TS$ decreases significantly with increasing $T$ and determines the $T$ dependence of $\Omega$.

\subsubsection{Estimation of $T_{\rm c}$}
Since $\Omega$ at lower and $\Omega$ at higher $T$ are dominated by the contribution from the internal energy $U_{\rm int}$ and the entropy $-TS$, respectively, the magnitude of $T_{\rm c}$ is estimated by extracting the entropic gain of the system. To do this, (i) we first calculate the $T$ dependence of the free energy $\Omega_{\rm FP}(T)$ for free particles system, in which there is no attractive Coulomb interaction between the electron and hole; (ii) Next, we compute the free energy difference $\Delta E \equiv \Omega_{\rm FP}(T) - \Omega(T)$ in the limit $T \rightarrow \infty$. This is the energy gain due to the attractive Coulomb interaction when charges are completely delocalized over the system; and (iii) finally, we compare $\Omega(T)$ in Eq.~(\ref{eq:OmegaTB}) with the free energy $\Omega_{\rm FP}(T) - \Delta E$. Here we subtracted the energy $\Delta E$ from $\Omega_{\rm FP}(T)$ to take into account the attractive Coulomb interaction effectively in the free particle system, which allows us to study the localization-delocalization transition as a crossover of the free energy. Figure \ref{fig:free_energy}(a) shows the $T$ dependence of $\Omega_{\rm FP}(T)$ (green dot-dashed), $\Omega_{\rm FP}(T)-\Delta E$ (blue dashed), and $\Omega(T=0)$ (black dotted) as well as $\Omega(T)$. Here, the magnitude of $\Delta E$ was estimated to $1.3t_0$. $\Omega_{\rm FP}(T)$ and $\Omega_{\rm FP}(T)-\Delta E$ decrease monotonically as $T$ increases. This is because there is no isolated peak in the EH DOS in the free particle system, similarly to the case of Fig.~\ref{fig:dos}(b). At finite $T$, the EH pair accommodates higher levels as well as the lowest energy level, which enhances the entropic gain of the system significantly. As $T$ increases from zero, the free energy difference between $\Omega$ and $\Omega_{\rm FP}-\Delta E$ becomes small. When the energy difference is smaller than zero, $\Omega$ starts to decrease, which determines $T_{\rm c}$. The magnitude of $T_{\rm c}$ is approximately given by the temperature at a crossing point (arrow) of the free energy curve $\Omega_{\rm FP}(T)-\Delta E$ and the straight line $\Omega(T=0)$ (see Fig.~\ref{fig:free_energy}).  Hence the approximate relation for $T_{\rm c}$ is given by  
\begin{eqnarray}
 \Omega_{\rm FP}(T_{\rm c}) - \Omega(T=0) - \Delta E = 0.
\label{eq:formula}
\end{eqnarray}
This relation may be useful for deriving $T_{\rm c}$ when we consider the long-range and anisotropic hopping effects below, while the transition temperature estimated by Eq.~(\ref{eq:formula}), i.e., $k_{\rm B}T/t_0\simeq$0.24 (arrow), somewhat underestimates the magnitude of $k_{\rm B}T_{\rm c}/t_0\simeq$0.27.


\subsubsection{Delocalized carrier (quantum mechanical) treatment}
We emphasize the importance of the quantum mechanical description of the carriers. In the present TB model, the electron and hole are treated by delocalized carriers by solving Eq.~(\ref{eq:sch}) self-consistently at each $T$. As discussed, the increase in $E_{\alpha}^{\rm (eh)}$ with increasing $T$ shown in Fig.~\ref{fig:Eeh}, which plays a key role in the localization-delocalization transition, is purely originated from the delocalized carrier treatment. If we treat the electron and/or hole as a strongly localized particle like a classical particle, it is very difficult to observe the CTE dissociation. This is easily seen if we consider, for example, the following potential
\begin{eqnarray}
 w_{\bm{p}}^{({\rm e})} &=&
 \begin{cases}
  w_0 & p_z \le -1 \\
 0 & p_z\ge 0 
 \end{cases},
 \nonumber\\
  w_{\bm{p}}^{({\rm h})} &=&
 \begin{cases}
  -U_1 & \bm{p}=(0,0,-1) \\
 w_0 & p_z \ge 0 \\
 0 & {\rm otherwise} 
 \end{cases},
 \label{eq:trap}
\end{eqnarray}
where $U_1 (\ge U_0)$ is a depth of the trapped potential for a hole. The solution of Eq.~(\ref{eq:sch}) with the use of this potential yields the CTE, where a hole is localized at $\bm{p}=(0,0,-1)$ (see also Fig.~\ref{fig:DA_TB}). The hole is strongly trapped by both the potential of $-U_1$ in Eq.~(\ref{eq:trap}) and the attractive Coulomb interaction with the electron. No delocalization of the hole enhances the attractive Coulomb interaction energy between the electron at $\bm{p}=(0,0,0)$ and the hole at $\bm{p}=(0,0,-1)$ through the second term in Eq.~(\ref{eq:Vele}). Simultaneously, the interaction energy between them is enhanced through the second term in Eq.~(\ref{eq:Vhole}), yielding the strongly bounded CTE at the DA interface. The value of $E_{\alpha}^{\rm (eh)}$ would increase very moderately compared with the case in Fig.~\ref{fig:Eeh}. Furthermore, the free energy $\Omega(T=0)$ of the system with such a CTE is much lower than $\Omega_{\rm FP}(T=0)$. Such a situation would lead to no localization-delocalization transition at a realistic $T$. 

\begin{figure*}[t]
\center
\includegraphics[scale=0.8,clip]{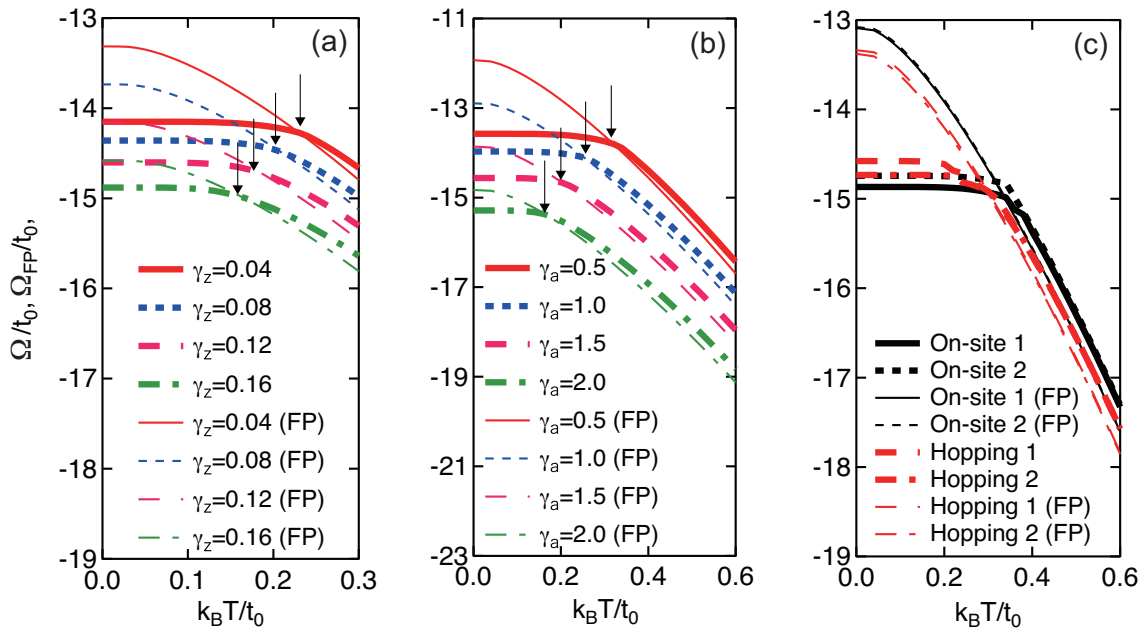}
\caption{\label{fig:hopping} (Color online) The $T$ dependence of the free energies $\Omega(T)$ (thick curves) given by Eq.~(\ref{eq:OmegaTB}) and $\Omega_{\rm FP}(T)-\Delta E$ (thin curves) for various (a) $\gamma_z$ and (b) $\gamma_{\rm a}$. The critical temperature $T_{\rm c}$ is indicated by an arrow as a crossing point between them. (c) $\Omega(T)$ (thick curves) and $\Omega_{\rm FP}(T)-\Delta E$ (thin curves) are plotted as a function of $T$ for the disordered systems with on-site random potentials and random hopping integrals. The curves for typical two samples are shown.}
\end{figure*}

\subsection{Hopping dependence}
\label{sec:hopping}
In Sec.~\ref{sec:mechanism}, we have considered a simple TB model with the nearest-neighbor hopping integral only and shown that there is a critical temperature $T_{\rm c}$ at which the CTE starts to dissociate. We next investigate how the long-range and anisotropic hopping affect the magnitude of $T_{\rm c}$. To study the effect of the long-range hopping, we assume the following parameters
\begin{eqnarray}
 t_{\bm{p},\bm{p}+\bm{R}_{1i}}^{({\rm e})} &=&
 t_{\bm{p},\bm{p}+\bm{R}_{1i}}^{({\rm h})} = t_0 \ (i = x, y, z), \nonumber\\
 t_{\bm{p},\bm{p}+\bm{R}_{2z}}^{({\rm e})} &=&
 t_{\bm{p},\bm{p}+\bm{R}_{2z}}^{({\rm h})} = \gamma_z t_0,
\end{eqnarray}
where $\gamma_zt_0$ is the next nearest-neighbor hopping parameter for $z$-direction. We also consider the following situation to study the effect of the anisotropic hopping, 
\begin{eqnarray}
 t_{\bm{p},\bm{p}+\bm{R}_{1i}}^{({\rm e})} = t_0, \ \ 
 t_{\bm{p},\bm{p}+\bm{R}_{1i}}^{({\rm h})} = \gamma_{\rm a}t_0 \ (i = x, y, z),
\end{eqnarray}
where $\gamma_{\rm a}t_0$ is the nearest-neighbor hopping parameter for a hole.

Figures \ref{fig:hopping}(a) and \ref{fig:hopping}(b) show the $T$ dependence of $\Omega$ and $\Omega_{\rm FP}-\Delta E$ for various $\gamma_z$ and $\gamma_{\rm a}$, respectively. The values of $U_0/t_0$, $w_0/t_0$, and $\Delta E/t_0$ are set to 10, 10, and 1.3, respectively. Similar to Sec.~\ref{sec:mechanism}, the free energy anomalies are observed at critical $T$s (arrows). As $\gamma_z$ and $\gamma_{\rm a}$ increases, $T_{\rm c}$ decreases monotonically. This behavior can be understood by Eq.~(\ref{eq:formula}). In general, for larger $\gamma_z$ and $\gamma_{\rm a}$, the band width for free particle system is large. Then, the lowest eigenenergy (and the band energy) decreases, which lowers the magnitude of $\Omega_{\rm FP}(T =0) - \Omega(T=0)$ since the value of $\Omega(T=0)$ is less sensitive to $\gamma_z$ and $\gamma_{\rm a}$. The small difference between the free energies at $T=0$ is effective to satisfy the relation of Eq.~(\ref{eq:formula}) at lower $T$. 

It should be noted that the large value of $\gamma_z$ and $\gamma_{\rm a}$ correspond to the light carrier mass. Several works have pointed out the importance of the light effective mass on the CTE dissociation \cite{arkhipov,wiemer,schwarz,ono}. The present work also supports this conclusion.

\subsection{Effect of disorder}
\label{sec:disorder}
It has been suggested that the disorder can assist the CTE dissociation \cite{peumans,rubel,nenashev,raos}. On the other hand, it has been also reported that the disorder makes the charge transfer ineffective \cite{li}. To investigate the disorder effect in the present model, we study two cases: the presence of the on-site random potentials and the random hopping integrals. The former case is modeled as 
\begin{eqnarray}
 w_{\bm{p}}^{({\rm e})} = 
 \begin{cases}
  w_0 + W_{\bm{p}} & p_z \le -1 \\
 W_{\bm{p}} & p_z\ge 0 
 \end{cases},
\nonumber\\
  w_{\bm{p}}^{({\rm h})} = 
 \begin{cases}
 -W_{\bm{p}}  & p_z \le -1 \\
 w_0 -W_{\bm{p}} & p_z \ge 0
 \end{cases},
 \label{eq:rand_pot}
\end{eqnarray}
where $W_{\bm{p}}/t_0 \in [-Z,Z]$ with a positive $Z$ that determines the strength of the random potential at site $\bm{p}$, while the latter case is modeled as
\begin{eqnarray}
 t_{\bm{p},\bm{p}+\bm{R}_{1i}}^{({\rm e})} &=&
 t_{\bm{p},\bm{p}+\bm{R}_{1i}}^{({\rm h})} = t_0 + W_{\bm{p}+\bm{R}_{1i}} \ (i = x, y, z).
 \nonumber\\
 \label{eq:rand_hop}
\end{eqnarray}
The values of $Z$, $U_0/t_0$, and $w_0/t_0$ are set to 1, 10, and 10, respectively.
Figure \ref{fig:hopping}(c) shows the $T$ dependence of $\Omega$ and $\Omega_{\rm FP}-\Delta E$ ($\Delta E = 1.3t_0$) for disordered systems with on-site random potentials (sample 1 and 2) and random hopping parameters (sample 1 and 2) represented by Eqs.~(\ref{eq:rand_pot}) and (\ref{eq:rand_hop}), respectively. The value of $\Omega (T=0)$ decreases by about $t_0$ due to the presence of the disorder, while that of $\Omega_{\rm FP} (T=0)$ is almost unchanged [see Fig.~\ref{fig:free_energy}(a)]. Since the free energy difference between them at $T=0$ is large, the transition temperature becomes large ($k_{\rm B}T_{\rm c}/t_0\ge$0.3). This indicates that the disorder may not assist the CTE dissociation. It may be because the disorder given by Eqs.~(\ref{eq:rand_pot}) and (\ref{eq:rand_hop}) confines the carrier distribution to the sites just near the DA interface, which lowers the free energy due to the attractive Coulomb interaction. Unless we consider different types of disorder, we would not observe the decrease in the $T_{\rm c}$.

\begin{figure*}[t]
\center
\includegraphics[scale=0.5,clip]{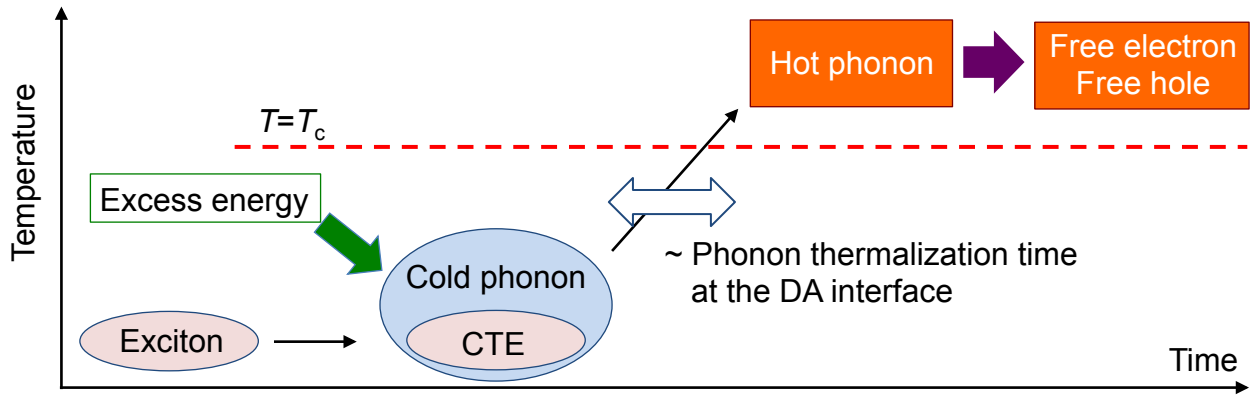}
\caption{\label{fig:scenario} (Color online) Temperature evolution of the charge carriers and phonons, provided that both the entropy mechanism proposed in the present study and the excess energy mechanism \cite{jackson} hold. The excess energy created by the CTE formation excites the phonon at the DA interface. The phonon temperature increases with time and becomes over $T_{\rm c}$ within the phonon thermalization time. Above $T_{\rm c}$, the CTE dissociates into the free electron and hole.}
\end{figure*}

\section{Discussion}
\label{sec:discussion}
The present work supports the idea of the entropy-driven CTE dissociation \cite{clarke,gregg,gao,monahan}. The combination with the idea of carrier delocalization \cite{deibel,nenashev,raos,ono} is crucial. The important findings in this work are (i) the existence of a critical temperature $T_{\rm c}$, below and above which the localization and delocalization of charge carriers are observed, respectively, and (ii) the approximate formula for determining $T_{\rm c}$ given by Eq.~(\ref{eq:formula}). The former is significant to discuss the origin of the CTE dissociation in Refs.~\cite{gao,monahan}, while the latter is useful to predict the magnitude of $T_{\rm c}$ theoretically.

We estimate the magnitude of $T_{\rm c}$ in typical organic solar cells. We set $U_0\sim$0.5 eV by using $\varepsilon \simeq 3 \varepsilon_0$ ($\varepsilon_0$ is the dielectric constant of vacuum) and the equilibrium molecule-molecule distance $d \simeq 1$ nm of C$_{60}$ crystals. The hopping parameter at the DA interface is set to $t_0 \simeq U_0/10$, by assuming that the single-particle band width is a few hundred meV. The height of the barrier potential is set to $w_0 \simeq U_0$, so that the CTE (not the Frenkel exciton) is formed at $T=0$. Then, the value of $k_{\rm B}T_{\rm c}/t_0 \simeq $0.27 corresponds to $T\simeq$160 K. The magnitude of this temperature is in agreement with the temperature, at which the free carrier density starts to increase, as observed in C$_{60}$-related organic solar cells \cite{gao}. 

It is noteworthy that the magnitude of the CTE binding energy $E_{\rm B}$ can also be estimated from the EH DOS. Assuming that the continuum states start from $\alpha \simeq 10$ at lower $T$ shown in Fig.~\ref{fig:dos}(a), the $\nu$th CTE binding energy is $E_{\rm B}(\nu) = E_{\alpha=10}^{\rm (eh)} - E_{\alpha=\nu}^{\rm (eh)}$. For example, $E_{\rm B}(\nu=1) = 5.3t_0 \simeq$ 0.26 eV, which is an order of magnitude higher than thermal energy at room temperature, but is consistent with experimental observations \cite{morteani}.

We do not exclude the possibility that the CTE dissociation occurs even when the lattice temperature $T_0$ is initially below $T_{\rm c}$. If the excess energy from the offset in energy levels at the DA interface \cite{jackson} excites the phonon modes and enhance the phonon temperature with some delay time, the CTE dissociation can occur. A possible scenario is as follows (see Fig.~\ref{fig:scenario}): 
\begin{enumerate}
\item The exciton is initially created at the donor region by photon absorption.
\item The electron transfer occurs at the DA interface, yielding the CTE formation.
\item The excess energy created by the CTE formation (i.e., the energy difference between the donor LUMO and acceptor LUMO) excites phonons at the interface and disturbs the cold phonon distribution initially at $T_0$.
\item Through the phonon-phonon and phonon-electron scatterings, the phonon modes will obey the Bose distribution function with temperature $T'$ higher than $T_0$ after the phonon thermalization time.
\item When $T'$ is larger than $T_{\rm c}$, the CTE can dissociate.
\end{enumerate}
If this scenario holds, the magnitude of $T'$ gradually increases with time. Then, the CTE energy also increases with time, as expected by the $T$-dependent $E_{\alpha}^{({\rm eh})}$ shown in Fig.~\ref{fig:Eeh}(a). This behavior is quite similar to the experimental observations, where the CTE spontaneously climbs up the Coulomb potential at the pentacene-vacuum interface \cite{monahan}. For deeper understanding, it is necessarily to study the time-dependence of the interface phonon temperature $T'$. This may be studied in the framework of the nonequilibrium theory of phonons \cite{kabanov,ono2}. 

The estimation of $T_{\rm c}$ in realistic systems \cite{dowgiallo,ferguson,bassler} would be important in understanding the difference of the CTE properties in organic solar cells. We propose a method how to calculate each term in Eq.~(\ref{eq:formula}). First, $\Omega(T=0)$ is equivalent to the total energy of the system at zero $T$ and can be calculated accurately by using the DFT approach and assuming that one electron is excited into an unoccupied state leaving a hole behind. Second, $\Omega_{\rm FP}(T)$ may be calculated in the framework of the finite-$T$ DFT by ignoring the attractive interaction between an electron and a hole. Finally, $\Delta E$, defined as $\Omega_{\rm FP}(T) - \Omega(T)$ in the limit of $T\rightarrow \infty$, is estimated by calculating the attractive Coulomb interaction energy between the electron and hole. Such an EH pair density is obtained by the computation of $\Omega_{\rm FP}(T\rightarrow \infty)$. 

We also emphasize the finite-$T$ effect on the excitonic properties. The exciton is usually described within many-body perturbation theory or time-dependent DFT \cite{onida}. Recently, the CTE has been studied in such a first-principles context \cite{cudazzo,petrone}. The extension to the $T$-dependent Bethe-Salpeter or time-dependent Kohn-Sham equations and their solutions would give an accurate estimation of the CTE binding energy and predict the localization-delocalization transition or the free energy anomaly mentioned in the present work. 

In the present TB model, we assumed that the dielectric constant is uniform across the DA interface [see Eqs.~(\ref{eq:Vele}) and (\ref{eq:Vhole})]. However, it may be natural to consider that the bonding arrangement between molecules is not uniform near the DA interface. In this case, the dielectric constant varies locally \cite{giustino2}. As a first approximation, the Coulomb interaction between the electron at $\bm{p}_{1}$ and the hole at $\bm{p}_2$ can be expressed as
\begin{eqnarray}
 -\frac{e^2}{4\pi d\sqrt{\varepsilon (\bm{p}_1) \varepsilon (\bm{p}_2)}} \frac{1}{\vert \bm{p}_1 - \bm{p}_2\vert},
 \label{eq:local}
\end{eqnarray}
where $\varepsilon (\bm{p}_i)$ with $i=1,2$ is the local dielectric constant at the site $\bm{p}_i$ \cite{ono}. It would be interesting to investigate the effect of the local dielectric constant [the use of Eq.~(\ref{eq:local}) in Eqs.~(\ref{eq:Vele}) and (\ref{eq:Vhole})] on the CTE dissociation.

\section{Summary}
\label{sec:Conc}
We have derived the $T$-dependent TB model for a EH pair at the DA interface, which enabled us to study the entropy as well as the carrier delocalization effect on the CTE dissociation. Our numerical calculations have revealed that there exists the localization-delocalization transition at a critical temperature $T_{\rm c}$, above which the CTE dissociates. This is related to the anomaly of the free energy $\Omega$. Below and above $T_{\rm c}$, $\Omega$ is determined by the internal energy and the entropic energy, respectively. The transition can be observed only when the carrier delocalization treatment is employed. Thus, the CTE dissociation has been successfully explained by a combination of quantum mechanics and thermodynamics. Analysis of the $T$ dependence of $\Omega$ has enabled us to derive the useful relation for $T_{\rm c}$ given in Eq.~(\ref{eq:formula}). The magnitude of $T_{\rm c}$ and the CTE binding energy estimated were in agreement with the experimental data. A possible scenario involving the phonon thermalization has been discussed.

So far, several origins for the CTE dissociation have been proposed (see Table~\ref{tab:previous}). The present work emphasized the importance of the combined effect of both the entropy (finite-$T$) and the carrier delocalization. The present study agrees with the notion that the light effective mass is effective in the CTE dissociation, but disagrees with the notion that the disorder effect is important. Our work would be the first step for understanding the entropy effect on the CTE dissociation observed at various DA interface in a unified manner. We hope that the localization-delocalization transition is observed in future experiments.


\begin{acknowledgments}
This study is supported by a Grant-in-Aid for Young Scientists B (No. 15K17435) from JSPS.
\end{acknowledgments}



\begin{thebibliography}{99}

\bibitem{few} S. Few, J. M. Frost, and J. Nelson, Models of charge pair generation in organic solar cells, Phys. Chem. Chem. Phys. {\bf 17}, 2311 (2015).

\bibitem{arkhipov} V. I. Arkhipov, P. Heremans, and H. B\"{a}ssler, Why is exciton dissociation so efficient at the interface between a conjugated polymer and an electron acceptor?, Appl. Phys. Lett. {\bf 82}, 4605 (2003).

\bibitem{wiemer} M. Wiemer, A. V. Nenashev, F. Jansson, and S. D. Baranovskii, On the efficiency of exciton dissociation at the interface between a conjugated polymer and an electron acceptor, Appl. Phys. Lett. {\bf 99}, 013302 (2011).

\bibitem{peumans} P. Peumans and S. R. Forrest, Separation of geminate charge-pairs at donor-acceptor interfaces in disordered solids, Chem. Phys. Lett. {\bf 398}, 27 (2004).

\bibitem{rubel} O. Rubel, S. D. Baranovskii, W. Stolz, and F. Gebhard, Exact Solution for Hopping Dissociation of Geminate Electron-Hole Pairs in a Disordered Chain, Phys. Rev. Lett. {\bf 100}, 196602 (2008).

\bibitem{deibel} C. Deibel, T. Strobel, and V. Dyakonov, Origin of the efficient polaron-pair dissociation in polymer-fullerene blends, Phys. Rev. Lett. {\bf 103}, 036402 (2009).

\bibitem{nenashev} A. V. Nenashev, S. D. Baranovskii, M. Wiemer, F. Jansson, R. \"{O}sterbacka, A. V. Dvurechenskii, and F. Gebhard, Theory of exciton dissociation at the interface between a conjugated polymer and an electron acceptor, Phys. Rev. B {\bf 84}, 035210 (2011).

\bibitem{schwarz} C. Schwarz, S. Tscheuschner, J. Frisch, S. Winkler, N. Koch, H. B\"{a}ssler, and A. K\"{o}hler, Role of the effective mass and interfacial dipoles on exciton dissociation in organic donor-acceptor solar cells, Phys. Rev. B {\bf 87}, 155205 (2013).

\bibitem{clarke} T. M. Clarke and J. R. Durrant, Charge photogeneration in organic solar cells, Chem. Rev. {\bf 110}, 6736 (2010).

\bibitem{gregg} B. A. Gregg, Entropy of charge separation in organic photovoltaic cells: The benefit of higher dimensionality, J. Phys. Chem. Lett. {\bf 2}, 3013 (2011).

\bibitem{gao} F. Gao, W. Tress, J. Wang, and O. Ingan\"{a}s, Temperature dependence of charge carrier generation in organic photovoltaics, Phys. Rev. Lett. {\bf 114}, 128701 (2015).

\bibitem{monahan} N. R. Monahan, K. W. Williams, B. Kumar, C. Nuckolls, and X.-Y. Zhu, Direct observation of entropy-driven electron-hole separation at an organic semiconductor interface, Phys. Rev. Lett. {\bf 114}, 247003 (2015).

\bibitem{raos} G. Raos, M. Casalegno, and J. Id\'{e}, An effective two-orbital quantum chemical model for organic photovoltaic materials, J. Chem. Theory Comput. {\bf 10}, 364 (2014).

\bibitem{ono} S. Ono and K. Ohno, Minimal model for charge transfer excitons at the dielectric interface, Phys. Rev. B {\bf 93}, 121301(R) (2016).

\bibitem{miller} A. Miller and E. Abrahams, Impurity Conduction at Low Concentrations, Phys. Rev. {\bf 120}, 745 (1960).

\bibitem{mermin} N. D. Mermin, Thermal Properties of the Inhomogeneous Electron Gas, Phys. Rev. {\bf 137}, A1441 (1965).

\bibitem{chakra} B. Chakraborty and R. W. Siegel, Electron and positron response to atomic defects in solids: A theoretical study of the monovacancy and divacancy in aluminum, Phys. Rev. B {\bf 27}, 4535 (1983).

\bibitem{boronski} E. Boro\'{n}ski and R. M. Nieminen, Electron-positron density-functional theory, Phys. Rev. B {\bf 34}, 3820 (1986).

\bibitem{ono_ohno} S. Ono and K. Ohno, unpublished.



\bibitem{li} G. Li, A. Nitzan, and M. A. Ratner, Yield of exciton dissociation in a donor-acceptor photovoltaic junction, Phys. Chem. Chem. Phys. {\bf 14}, 14270 (2012).

\bibitem{morteani} A. C. Morteani, P. Sreearunothai, L. M. Herz, R. H. Friend, and C. Silva, Exciton Regeneration at Polymeric Semiconductor Heterojunctions, Phys. Rev. Lett. {\bf 92}, 247402 (2004).

\bibitem{jackson} N. E. Jackson, B. M. Savoie, T. J. Marks, L. X. Chen, and M. A. Ratner, The Next Breakthrough for Organic Photovoltaics?, J. Phys. Chem. Lett. {\rm 6}, 77 (2015).

\bibitem{kabanov} V. V. Kabanov, J. Demsar, B. Podobnik, and D. Mihailovic, Quasiparticle relaxation dynamics in superconductors with different gap structures: Theory and experiments on YBa$_2$Cu$_3$O$_{7-\delta}$, Phys. Rev. B {\bf 59}, 1497 (1999).

\bibitem{ono2} S. Ono, H. Shima, and Y. Toda, Theory of photoexcited carrier relaxation across the energy gap of phase-ordered materials, Phys. Rev. B {\bf 86}, 104512 (2012).


\bibitem{dowgiallo} A.-M. Dowgiallo, K. S. Mistry, J. C. Johnson, and J. L. Blackburn, Ultrafast spectroscopic signature of charge transfer between single-walled carbon nanotubes and C$_{60}$, ACS Nano {\bf 8}, 8573 (2014).

\bibitem{ferguson} A. J. Ferguson, A.-M. Dowgiallo, D. J. Bindl, K. S. Mistry, O. G. Reid, N. Kopidakis, M. S. Arnold, and J. L. Blackburn, Trap-limited carrier recombination in single-walled carbon nanotube heterojunctions with fullerene acceptor layers, Phys. Rev. B {\bf 91}, 245311 (2015).

\bibitem{bassler} H. B\"{a}ssler and A. K\"{o}hler, "Hot or cold": how do charge transfer states at the donor-acceptor interface of an organic solar cell dissociate?, Phys. Chem. Chem. Phys. {\bf 17}, 28451 (2015).

\bibitem{onida} G. Onida, L. Reining, and A. Rubio, Electronic excitations: density-functional versus many-body Green's-function approaches, Rev. Mod. Phys. {\bf 74}, 601 (2002).



\bibitem{cudazzo} P. Cudazzo, F. Sottile, A. Rubio, and M. Gatti, Exciton dispersion in molecular solids, J. Phys.: Condens. Matter {\bf 27}, 113204 (2015).

\bibitem{petrone} A. Petrone, D. B. Lingerfelt, N. Rega, and X. Li, From charge-transfer to a charge-separated state: a perspective from the real-time TDDFT excitonic dynamics, Phys. Chem. Chem. Phys. {\bf 16}, 24457 (2014). 

\bibitem{giustino2} F. Giustino and A. Pasquarello, Theory of atomic-scale dielectric permittivity at insulator interfaces, Phys. Rev. B {\bf 71}, 144104 (2005).












\end{thebibliography}
\end{document}